\documentclass[aps,showpacs,showkeys]{revtex4}
\usepackage{amsfonts,amsmath}

\begin{document}

\title{Fluctuating Commutative Geometry}
\author{Luiz C. de Albuquerque$^1$, Jorge L. deLyra$^2$, 
and Paulo Teotonio-Sobrinho$^2$}
\affiliation{(1) Faculdade de Tecnologia de S\~ao Paulo - DEG - CEETEPS - 
UNESP, Pra\c{c}a Fernando Prestes, 30, 01124-060 S\~ao Paulo, SP, Brazil}
\affiliation{(2) Universidade de S\~ao Paulo, Instituto de F\'{\i}sica - DFMA\\
Caixa Postal 66318, 05315-970, S\~ao Paulo, SP, Brazil}

\begin{abstract}
We use the framework of noncommutative geometry to define a discrete 
model for fluctuating geometry. Instead of considering ordinary
geometry and its metric fluctuations,  we consider
generalized geometries where topology and dimension can also
fluctuate. The model describes the geometry of
spaces with a countable number $n$ of points. The spectral principle
of Connes and Chamseddine is used to define dynamics.
We show that this simple model has two  phases. The expectation value 
$\langle n \rangle$, the average number of points in the universe,  
is finite in one phase and diverges in the other. Moreover, the  
dimension $\delta$ is a dynamical observable in our
model, and plays the role of an order parameter. The computation
of $\langle\delta\rangle$ is discussed and an upper bound is found, 
$\langle\delta\rangle\,<\,2$. We also address another discrete model
defined on a fixed $d=1$ dimension, where topology fluctuates.
We comment on a possible spontaneous localization of topology.
\end{abstract}

\maketitle
\newpage

\section{Introduction}

A possible approach to quantum gravity is to consider it as an Euclidean 
quantum field theory. In the functional integral approach, one tries to 
compute a sum over metrics $g_{\mu \nu}$ defined on a $d$-dimensional manifold 
$M$. The dynamics is fixed by a choice of an action that is usually
taken to be the Einstein-Hilbert action. 
An exact computation of such functional integral 
is not available and several approximated methods have been 
developed. The main idea is to use some 
kind of discretization in order to approximate the functional integral by a 
large number of ordinary integrals or sums. One such method is known
as the dynamical triangulation  and consist in summing over
the triangulations of a given manifold. The situation becomes 
more complicated when one wishes to include quantum topology 
change. In this case, the manifold itself is also not fixed and a  
sum over topologically non equivalent manifolds $M$ has to be 
included. The overall  picture seems to be under control only for $d=2$ 
where a classification of topologies is possible. Such a 
classification is proven to be impossible for  $d=4$ and it is an open
problem  for $d=3$. 

The framework of noncommutative geometry is appropriate
to explore some difficult questions in quantum gravity. We illustrate
how noncommutative geometry can be used to generalize  Euclidean
quantum gravity, i.e, fluctuating geometry. 
Instead of considering ordinary geometry and its metric fluctuations, 
we consider the generalized geometries where besides the metric,
topology and dimension can also fluctuate.

The basic idea coming from noncommutative geometry \cite{NCG} is that one can describe
a Riemannian manifold $(M,g_{\mu \nu})$ in a purely algebraic way.
There is no loss of information if, instead of the data $(M,g_{\mu
\nu})$, one is given a triple $({\cal A},{\cal H}, D)$, where
${\cal A}$ is the C*-algebra  $C^0(M)$ of smooth functions on $M$, 
${\cal H}$ is the Hilbert space of $L^2$-spinors
on $M$, and ${\cal D}$ is the Dirac operator acting on ${\cal
H}$. From the Gelfand-Naimark theorem it is known that
the topological space $M$ can be reconstructed from 
the set $\hat{\cal A}$ of
irreducible representations of $C^0 (M)$.
Metric is also encoded, and the geodesic distance can
be computed from  ${\cal D}$.  Here 
we will consider only commutative spectral triples - this
is enough to go much beyond ordinary geometry. In particular one 
can treat all Hausdorff topological spaces in this way. Given a pair
$(M,g_{\mu \nu})$, one can promptly construct the corresponding
triple $(C^0 (M),L^2(M),{\cal D})$. However, not all commutative
spectral triples, or generalized geometries, come from a pair 
$(M,g_{\mu \nu})$. Nevertheless one can always associate a 
Hausdorff space $M=\hat {\cal A}$ to a commutative spectral triple, 
where $\hat {\cal A}$ denotes the set of irreducible representations
of ${\cal A}$. However, the space $M$ may not be a manifold. 
Once we trade the original Riemannian geometry for its corresponding
commutative triple we need a replacement for the Einstein-Hilbert 
action $S_{EH}$. The so-called spectral action 
of Chamseddine and Connes \cite{CC} is one possible candidate. 
It depends only on the eigenvalues of ${\cal D}$ (the spectral
principle) and contains  $S_{EH}$ as a dominant term. 
In this paper however we shall use another spectral action.

The spectral action can be 
written for any triple, regardless of whether it comes from a manifold 
$(M,g_{\mu \nu})$ or not. In the spectral geometry approach  it is 
conceivable to write the partition function
\begin{equation}\label{1}
Z=\sum_{x\in {\cal X}}e^{-S[x]},
\end{equation}
were the \lq\lq sum'' is over the set ${\cal X}$ of all possible
commutative spectral triples and $S$ depends on the spectrum of
${\cal D}$. It includes all Hausdorff spaces
and therefore all manifolds of all dimensions.

Apparently, there is no advantage in considering  
the partition function (\ref{1}) since it is by no means easier to 
compute. However, the algebraic approach provides us with a 
natural way of defining discrete approximations for the theory. For that 
it is enough to replace the algebra ${\cal A}$ by a finite dimensional 
algebra $A_n$. In this approach to discretization
there is no need to introduce a lattice or simplicial 
decomposition of the underlying space. The approximation of ${\cal A}$ 
by a finite algebra works even if the spectral triple does not come from 
a  manifold. In this sense, it gives us a generalization of ordinary 
discretizations \cite{discrete}. The simplest discrete model one can consider is a 
simplification of (\ref{1}), where instead of summing over the set ${\cal X}$ 
of commutative spectral triples, we take a subset $X\subset {\cal X}$. The 
set $X$ consists of points $x=({\cal A},{\cal H},D)$ where ${\cal A}$ 
is a commutative algebra with a countable spectrum $\hat A$, i.e., with 
a countable number of irreducible representations. Therefore the 
underlining space $M=\hat {\cal A}$ has a countable and possibly 
finite number of points. 
 
Most of the results discussed here were reported in \cite{ALT}.
In section 2 we reintroduce the discrete model of \cite{ALT}. 
In section 3 we show that this simple model has two phases. 
The expectation value of the number of points diverges in one phase
and it is finite in the other phase.
The definition of dimension in noncommutative geometry is recalled in
section 4. A estimate of the expectation value of the dimension is
discussed in section 5. In section 6 we briefly discuss another model
for random geometry where the dimension is fixed and equal to one, but
the topology fluctuates. Further results and details on this second
model will be reported elsewhere \cite{ATV}.

\section{Discrete model}

The exact computation of (1) is a major goal, not yet
accomplished. In this paper we  discretize  (\ref{1}) by sampling the
set ${\cal X}$ with finite commutative spectral triples.
We will think of it as a useful toy model, which seems to 
capture some of the main features of the full one, Eq. (1). 
For instance, the key role played by the eigenvalues of the Dirac 
(or Laplace) operator
in the spectral action approach was emphasized in \cite{Landi}. In
our model they are also the natural dynamical variables due to the
connection with random matrix theory. An important
ingredient of the model is that the number of points can
fluctuate.  Moreover, in our simple model the space-time dimension
is a dynamical  observable and its expectation value can be  computable 
from first principles.

Let us describe the ensemble $X\subset \cal{X}$ of geometries we 
will consider.
A point of $x\in X$ is a commutative spectral triple $x=({\cal
A},{\cal H}, D )$ where the commutative \mbox{C*-algebra} ${\cal
A}$ has a countable spectrum ${\hat {\cal A}}$. We divide $X$ into
subspaces $X_n$ consisting of triples $({\cal A}_n,{\cal H}_n, D
)$ such that ${\hat {\cal A}}_n$ has a fixed number $n$ of points.
From the Gelfand-Naimark theorem it follows that elements of
${\cal A}_n$ are the (possibly infinite) sequences
$a=(a_1,a_2,...,a_n)$, $a_j\in{\mathbb{C}}$. The Hilbert space
${\cal H}_n$ is given by vectors $v=(v_1,...,v_n)$ with norm
$||v||^2\equiv\sum_{i=1}^n v_i^2 <\infty$. The elements of ${\cal
A}$ are represented by diagonal matrices $ \hat
a=\mbox{diag}(a_1,...,a_n)$ acting on ${\cal H}_n$. Finally, the
operator $D$ is a $n\times n$ self-adjoint matrix. We will sample
the space $X$ by $X_1,X_2,...,X_N$ and eventually take the limit
$N\rightarrow \infty $.

\section{Dynamics}

Let $L$ be a length scale
such that the operator ${\cal D}$
given by ${\cal D}=D/L$ will be the analogue of the Dirac
operator. The Chamseddine--Connes action
depends on a cutoff function of the eigenvalues of $D/L$.
The cutoff function is zero for eigenvalues of $D$  greater than 
$L$ and one otherwise \cite{CC,Landi}.
In other words, the Boltzmann weight in Eq.(\ref{1}) 
would be one outside a compact region in the eigenvalue space, 
leading to a divergent partition function 
(see Eq. (\ref{21})). Let us consider a quadratic
action instead:
\begin{equation}\label{4}
S[x]={\rm Tr}\left(\frac{{\cal D}}{\Lambda}\right)^2\equiv 
\beta ~{\rm Tr}(D^2),
\end{equation}
where $\Lambda $ is the inverse of Planck's length $l_p$, and
$\beta = (l_p/L)^2$. Finally, we define the partition function
$Z_N(\beta )= \sum _{n=0}^{N}z_n(\beta )$
where
\begin{equation}\label{5}
z_n(\beta)=
\int [dD] 
e^{-\beta{\rm Tr}(D^2)}
\end{equation}
is the partition function restricted to $X_n$,
in other words,
an integral over all independent matrix elements
$D_{ij}$, where  $[dD]$ is the usual measure for $n\times n$ Hermitian
matrices \cite{Meh}. The partition function $z_n(\beta)$ 
defines the one-matrix Gaussian Unitary Ensemble \cite{Meh}. 
A straightforward computation gives
\begin{equation}
z_n(\beta)= 2^{\frac{n}{2}}(\frac{\pi }{2\beta})^{\frac{n^2}{2}}.
\end{equation}

The expectation values of an observable  
${\cal O}(D_{ij})$ restricted to $X_n$ and for the entire ensemble
are 
\begin{eqnarray} \label{pts1}
\langle{\cal O}\rangle_{n,\beta}&\equiv& 
{\int [dD] {\cal O}e^{-\beta
{\rm Tr}(D^2)}}/{z_n(\beta)}~~\mbox{and}\\
\label{pts2}
\langle{\cal O}\rangle(\beta )&\equiv&\sum_{n=1}^{N}
P(n,\beta )\langle{\cal O}\rangle_{n,\beta},
\end{eqnarray}
respectively, where the function 
\begin{equation}
P(n,\beta )=\frac{z_n(\beta)}{\sum _n
z_{n}(\beta)}
\end{equation} 
is interpreted as the probability of having a
universe with $n$ points. The simplest observable in our model is
$n$, the number of points in $\hat {\cal A}$. By definition, $n$
is constant in $X_n$, therefore $\langle n\rangle_{n,\beta}=n$. 
Thus we get
\begin{equation}\label{11}
\langle n\rangle(\beta)=\frac{\sum _n n 2^{\frac{n}{2}}
(\frac{\pi}{2\beta})^{\frac{n^2}{2}}} 
{\sum _n 2^{\frac{n}{2}}(\frac{\pi}{2\beta})^{\frac{n^2}{2}}}.
\end{equation}
The mean $\langle n\rangle$ (\lq\lq average number of points in the 
universe'')
is not a continuous function of $\beta$ at $\beta_c=\pi/2 $, signaling the
onset of a phase transition.
Besides straightforward numerical calculation,
there are other ways to show that the
sum (\ref{11}) converges for $\beta>\beta_c$  and diverges for
$\beta<\beta_c$. 

\section{Defining Dimension $\delta $}

For $\beta<\beta_c$ the relevant universes have $\langle n\rangle=\infty$ 
and  $\Delta n/\langle n\rangle=0$. For
a $\infty$-dimensional $D$ one can define the dimension $\delta $
of the space $\hat {\cal A}$ from the eigenvalues of $D$. Let
$\{\mu_0(D),\mu_1(D),... \}$ be the modules of the eigenvalues
(i.e. the singular values) of $D$ organized in an increasing
order. By the Weyl formula \cite{NCG}, the dimension $\delta $  is
related to the asymptotic behavior of the eigenvalues for large
$k$: $\mu_k(D)\approx k^{\frac{1}{\delta }}$. By definition
$\delta =0$ for finite dimensional spectral triples. We can argue
that $\langle\delta\rangle$ is of the form
\begin{equation} \label{19}
\langle\delta\rangle(\beta)=\left\{
          \begin{array}{ll}
          f(\beta) & \mbox{if $\beta<\beta_c$}, \\
          0 & \mbox{if $\beta>\beta_c$}.
          \end{array}
        \right.
\end{equation}
This follows from the fact that for $\beta >\beta _c$ the
probability $P(n,\beta )$ is localized  around some finite $n$.
Hence $\langle\delta\rangle$ works as an order parameter. The value
$\beta_c=\pi/2$ separates $\langle\delta\rangle=0$ from the rest.

In order to study the  dimension we need to consider the
spectral $\zeta$-function

\begin{equation}\label{20}
\zeta(z)=\lim_{n\to\infty}\,
\sum_{k=0}^{n} \mu_k^{-z}={\rm Tr}\,(|D|^{-z}),
\end{equation}
where $D$ is an $\infty$-dimensional matrix ($\mu_0>0$). 
The relation between the dimension and $\zeta(z)$ has been 
discussed in \cite{Connes}. For large
enough values of $\alpha={\rm Re}(z)$, 
${\rm Tr}(|D|^{-z})$ is well defined. One says that $D$
has dimension spectrum $Sd$ if a discrete subset 
$Sd=\{s_1,s_2,...\}\subset{\mathbb{C}}$ exists, such that
$\zeta(z)$ can be holomorphically extended to ${\mathbb{C}}/Sd$.
This definition is consistent with the Weyl formula.
The set $Sd$ has more than a single point when for example
the geometry is the union of pieces of different dimensions
\cite{Connes}. In what follows we will look at an upper bound for
the dimension: It may happens that ${\rm Tr}(|D|^{-\alpha})=0$
for large enough $\alpha$, whereas
for small values of $\alpha$,  ${\rm Tr}(|D|^{-\alpha})= \infty$. 
Eventually, there is a value of
$\alpha$ (say, $\alpha_c$) for which ${\rm
Tr}(|D|^{-\alpha_c})$ is finite and non-zero. 
The upper bound for the  dimension will be $\delta=\alpha_c$. 

\section{Computing $\langle\delta\rangle$}

In order to estimate $\langle\delta\rangle$ by means of (\ref{20}), 
we rewrite (\ref{5}) and
(\ref{pts1}) as integrals over the eigenvalues $\lambda_k$ of $D$.
The procedure is well-known \cite{Meh}, and leads to
($C_n\equiv \pi^{\frac{n(n-1)}{2}}/\prod_{k=1}^n k!$)
\begin{equation}\label{21}
z_n(\beta)=C_n
\int_{-\infty}^\infty [d^n\lambda]\Delta^2(\lambda_k)
e^{-\beta\sum_{i=1}^n \lambda_i^2} \equiv 
C_n\Psi_{n,\beta}~,
\end{equation}

\begin{eqnarray}
\langle{\cal O}(\lambda_i)\rangle_{n,\beta} &=&\int_{-\infty}^\infty
[d^n\lambda]{\cal O}(\lambda_i)\left\{\frac{2^{\frac{n(n-1)}{2}}
\beta^{\frac{n^2}{2}}}
{\pi^{\frac{n}{2}}\prod_{k=1}^n k!} \Delta^2(\lambda_k)
e^{-\beta\sum_{i=1}^n \lambda_i^2}\right\}\\
&\equiv&
\int_{-\infty}^\infty
 [d^n\lambda]{\cal O}(\lambda_i)
{\cal P}_{n,\beta}(\lambda_k)~,\label{23}
\end{eqnarray}
where $\Delta(\lambda_k)=\prod_{i<j}(\lambda_j-\lambda_i)$ is the
Vandermonde determinant (Jastrow factor), and $[d^n\lambda]\equiv
\prod_{i=1}^n d\lambda_i$.

In random matrix theory, 
$\Psi_{n,\beta}(\gamma)$ is interpreted as the positional
partition function of an ensemble of equal charged particles (with
positions given by $\lambda_i$) in two dimensions, moving along an
infinite line, in thermodynamic equilibrium at temperature
$\gamma$ - the so-called \lq\lq Dyson gas'' \cite{Dy}. 
Then, ${\cal P}_{n,\beta}(\lambda_1,...,\lambda_n)$
defined in (\ref{23}) is the probability of finding one particle
at $\lambda_1$, one at $\lambda_2$, etc.
The value of $\Psi_{n,\beta}(\gamma)$ is known from the Selberg's
integral. 

In the region $\beta\leq\beta_c$ the partition function $Z(\beta)$
is dominated by $\infty-$dimensional matrices. Thus, one may try
to select the $\infty-$dimensional matrices out of the whole
ensemble, and then compute the mean of the $\zeta$-functions
following (\ref{20}). However, from the standpoint of our
statistical approach this procedure does not seems natural since
the sum over $n$ is a key ingredient in the whole construction.
Hence, we look for a quantity related to the $\zeta$-function that
captures the statistical nature of our model. Let us compute the
mean value 
\begin{equation}\label{25}
\langle{\rm
Tr}_\kappa\,|D|^{-\alpha}\rangle_{n,\beta}\,=
\,\left\langle\sum_{k=1}^n\,|\lambda_k|^{-\alpha}\,
\theta(|\lambda_k|-\kappa)\right\rangle_{n,\beta}.
\end{equation}

After some considerations one arrives at \cite{ALT}
\begin{equation}\label{30}
\langle{\rm Tr}_{\kappa_{n,\beta}}\,|D|^{-\alpha}\rangle_{n,\beta}\,
\approx\,
\frac{2}{\pi}\,(2n)^{1-\frac{\alpha}{2}}\,\beta^{\frac{\alpha}{2}}\,
\int_{\epsilon}^1\,\frac{dy}{y^\alpha}\,\sqrt{1-y^2}.
\end{equation}
The asymptotic behavior of $\langle{\rm
Tr}_{\kappa_{n,\beta}}\,|D|^{-\alpha}\rangle_{n,\beta}$ does not depend
in an essential way on the particular choice of $\epsilon$, as
long as we keep $\epsilon\neq0$.

Now we use the asymptotic formula (\ref{30}) and
search for the value $\alpha_c$ for which, as $N\to\infty$ and
$\beta\to\beta_c$, $\langle{\rm
Tr}_{\kappa_{n,\beta}}\,|D|^{-\alpha}\rangle$ diverges (converges to
zero) if $\alpha<\alpha_c$ ($\alpha>\alpha_c$), with $\langle{\rm
Tr}_{\kappa_{n,\beta}}\,|D|^{-\alpha_c}\rangle$ finite and non-zero. 
This gives an upper bound for the dimension of the \lq\lq condensed'' 
manifold in the infinite phase ( $\beta\leq\beta_c$), which is
$\langle\delta\rangle< \alpha_c$. We obtain:

\begin{equation}\label{31} 
\langle{\rm Tr}_{\kappa_{n,\beta}}\,|D|^{-\alpha}
\rangle(\beta)\,\sim\,\lim_{N\to\infty}\,
\sum_{n=1}^{N}\, P(n,\beta)\,n^{1-\frac{\alpha}{2}}~.
\end{equation} 
In the finite phase ($\beta>\beta_c$) the sum
in (\ref{31}) converges for $\alpha\geq 0 $. We conclude 
that  $\alpha_c=0$
(i.e. $\langle\delta\rangle=0$) for $\beta>\beta_c$ , as expected.
From the behavior of $P(n,\beta)$ in the 
infinite phase it follows that the convergence  of
the sum in (\ref{31}) is dictated by the behavior of
$\Gamma_{n,\alpha}= n^{1-\frac{\alpha}{2}}$ in the
limit $n\sim N\to\infty$.  For
$\beta\leq\beta_c$ we get $\Gamma_{n,\alpha}\to\infty$ if
$\alpha<2$, and $\Gamma_{n,\alpha}\to0$ if $\alpha>2$. For
$\alpha=2$ it turns out that $\langle{\rm
Tr}_{\kappa_{n,\beta}}\,|D|^{-2}\rangle(\beta)\,\sim 1$.
Therefore, we obtain the upper bound 
$\langle\delta\rangle < 2$.

\section{One dimensional model}

We would like to apply the ideas of last sections to 
spaces of dimension one instead of dimension zero.
Let us consider a collection $X$ of $n$ one dimensional intervals. 
The corresponding spectral triple will be $(A_X,{\cal H}_X, D)$ where
$A_X$ is the
algebra of continuous functions on $X$ and ${\cal H}_X=L^2(X)$. The
analogue of the Dirac operator will be the momentum operator
$-i\frac{\partial}{\partial x}$. We will keep $X$ fixed and fluctuate
$D$. 

Let us consider a simple example where $X$ is a pair of
disjoint intervals $I_1,I_2$. The intervals will be 
parametrized  by coordinate $x\in [0,2\pi]$.
An element $\psi \in {\cal H}_X$ is a pair of functions 
$\psi _1(x),\psi _2(x)$, \mbox{$\psi _i:I_i\rightarrow \mathbb{C}$} and the 
scalar product is
\begin{equation}
(\psi , \chi )=\int _0^{2\pi}dx(\psi _1^*\chi _1 + 
                  \psi _2\chi _2)\label{2.8}
\end{equation}

We can write $\psi $ as a column vector and the operator $D$ 
as the following matrix
\begin{equation}
D=
\left(
\begin{array}{cc}
  -i\partial _x & 0\\
  0 & -i\partial _x 
\end{array}
\right)\label{2.9}
\end{equation}

We have not fixed completely the spectral triple. The operator $D$ is
fixed only up to boundary conditions (BC's) or self-adjoint
extensions. The most general BC can be written as
\begin{equation}
\left( \begin{array}{c}  
        \psi _1(2\pi) \\
          \psi _2(2\pi)      
       \end{array}
\right)
=
\left( \begin{array}{cc}  
        g_{11} & g_{12} \\
          g_{21} & g_{22}     
       \end{array}
\right)
\left( \begin{array}{c}  
        \psi _1(0) \\
          \psi _2(0)      
       \end{array}
\right),
\end{equation}
where $g$ is a matrix in $U(2)$. 

We may ask what geometrical properties of $X$ is determined
by such BC's.
The point of view taken by Balachandran at all in  \cite{Bal} is that a
BC fixes the global topology. A couple of examples will illustrate
their point of view. First let us suppose $g_{ij}=\delta_{ij}$. In this case 
$\psi_i(0)=\psi_i(2\pi)$. The two intervals fold into a pair of
independent circles. Now suppose $g_{11}=g_{22}=0$ and
$g_{12}=g_{21}=1$. Therefore $\psi_1(2\pi)=\psi_2(0)$ and 
$\psi_2(2\pi)=\psi_1(0)$. As one can see, the two intervals are
connected to make a single circle of size $4\pi$. For a
generic BC, however, topology is not localized as in these two examples but 
it is rather a superposition of both.
We refer to \cite{Bal} for further details.

We would like to compute the partition function
\begin{equation}
Z_X(\beta)=\sum _{D}e^{-\beta TrD^2}\label{2.1}
\end{equation}
and look at the probability distribution for the BC's. For finite
values of $\beta $ the 
topology, as given by the
BC, will fluctuate. We have evidence, however, that the BC gets
localized as we increase $\beta $ and becomes a pair of circles in the
infinite limit. The significance of this fact and generalizations of
this one dimensional model will appear in \cite{ATV}. 

\section{Final Remarks}

We proposed a discrete model for Euclidean quantum
gravity, i.e, random geometry,  
based on the framework of noncommutative geometry. 
The first model contains the mean number of points, 
$\langle n\rangle$,
and the dimension of the space-time, $\langle\delta\rangle$, as dynamical
observables. We have shown that the discrete model has two phases: a finite
phase with a finite value of $\langle n\rangle$ and 
$\langle\delta\rangle=0$, and an
infinite phase with a diverging $\langle n\rangle$ and a finite 
$\langle\delta\rangle\ne0$.
An upper bound 
for the order
parameter $\langle\delta\rangle$ was found, $\langle\delta\rangle < 2$. 
We also considered the simplest example of another model where 
dimension is fixed but topology fluctuates. In 
the limit of infinite $\beta $, however, topology gets localized .

\section*{Acknowledgments}
The work of L.C.A. was partially supported by
FAPESP, grant 00/03277-3.

\end{document}